\documentclass[a4paper]{jpconf}
\usepackage{graphicx}

\usepackage{aasjournals}
\usepackage{amsmath}
\usepackage{amssymb}
\usepackage{hyperref}
\usepackage[sort&compress,numbers]{natbib}

\usepackage{color}
\begin{document}

\title{Ookami: An A64FX Computing Resource}

\author{A~C~Calder$^{1,2}$, 
E~Siegmann$^{2}$, C~Feldman$^{1,2}$, S~Chheda$^2$, D~C~Smolarski$^3$, F~D~Swesty$^1$, A~Curtis$^2$, J~Dey$^2$, 
D~Carlson$^2$, B~Michalowicz$^4$,
and
R~J~Harrison$^2$}

\address{$^1$ Department of Physics and Astronomy, 
Stony Brook University, Stony Brook, NY 11794-3800, USA}
\address{$^2$ Institute for Advanced Computational Science,
Stony Brook University, Stony Brook, NY 11794-5250, USA}
\address{$^3$ Department of Mathematics \& Computer Science, Santa Clara University, Santa Clara, CA 95053, USA}
%Add CS dept at SBU?
\address{$^4$ Department of Computer Science and Engineering, The Ohio State University, Columbus, 0H 43210, USA}

\ead{alan.calder@stonybrook.edu}

\begin{abstract}
We present a look at Ookami, a project providing community access to a testbed supercomputer with
the ARM-based A64FX processors developed
by a collaboration between RIKEN and Fujitsu and
deployed in the Japanese supercomputer Fugaku.
%the World's first exascale supercomputer. 
% i don't think this is true ^^.  exascale acc. to linpack? - then that's frontier.  -smeet
We describe the project, provide details about the 
user base and education/training program, and 
present highlights from performance studies of
two astrophysical simulation codes. 
\end{abstract}

\section{Introduction}

In this paper, we describe Ookami, a project providing access
to a testbed computing system featuring the
ARM-based A64FX processor developed by a collaboration between RIKEN and Fujitsu and deployed
in Fugaku, what was, until June of 2022,
% 2022 - frontier sets exaflop performance mark with hpl - unless i'm mistaken about the year 
the world's fastest supercomputer \cite{fugaku_exa}. The project is
run by Stony Brook University (SBU) in cooperation with the University at Buffalo~\cite{ookamiurl} and 
provides open access along with training and resources to effectively use such hardware. 
Users have been able to port, analyze, and optimize the performance of many applications~\cite{pearc_experiences_2021}. 
Below, we describe the hardware, our user base and 
education program, and provide highlights from performance studies of two astrophysical simulation codes,
the astrophysical radiation hydrodynamics code V2D \cite{sm09} and multi-application package FLASH \cite{Fryxetal00,calder.fryxell.ea:on,flash_development}, here
applied to thermonuclear supernovae.

\section{Overview of Project}
%Description of scale of project. FTEs?
%Dave: Perhaps mention Buffalo collaboration here as well?
%Mentioned buffalo above. Think that will do it.

The primary goal of the Ookami project is to provide
access to a platform with the A64FX processors for testing
and development.  These processors offer an alternative 
to graphical programming units (GPUs) that power many contemporary
supercomputers but may require considerable code development 
to make good use of.  The expectation is that the reduced instruction set A64FX processors will provide high 
performance and reliability for applications with common programming 
models, particularly memory-intensive applications, while
maintaining a good performance-to-power ratio similar
to GPUs. Users with a variety of applications are
able to explore and evaluate this new hardware and its
potential for use in settings from a local cluster to
the extreme scale of leadership-class supercomputing centers. 

The Ookami project started in 2019 and will run for six years, providing researchers worldwide with access to this cutting-edge computing technology. For the first few project years, access to 
Ookami was allocated via SBU. Researchers submitted allocation requests directly to the Ookami team, who reviewed them. Approved projects were granted access to the cluster for testbed projects, and once users could prove that their 
application performs well on this novel architecture, they could advance their projects to production status. In October 2022 Ookami became an ACCESS \cite{nsfaccess} resource provider, and now 
90\% of its resources are allocated via ACCESS. Users can submit requests to ACCESS and, if approved, get credits, which they can then exchange for resources (e.g.\ node hours, GPU hours, storage) on one or multiple of the ACCESS resource providers. 

\subsection{The A64FX Hardware}

Ookami is an HPE (formerly Cray) Apollo 80 system with
174 + 2 debug
A64FX-FX700 Fujitsu compute nodes,
each of which
has 48 cores divided into 4 core memory groups (NUMA regions), each with 8GB of high-bandwidth memory (HBM) for a total
of 32 GB per node. 
Each core has a 64 KB L1 cache, and an 8 MB L2 cache shared between the cores in each core memory group, and runs at 1.8 GHz.
These processors use the ARMv8.2--A
Scalable Vector Extension (SVE) SIMD instruction set with a 512-bit vector
implementation, allowing for vector lengths anywhere from 128 to 2048 bits
(in 128-bit increments)
and enabling vector length agnostic programming~\cite{armv8.2}. 

The  operating  system for each
processor resides  on a node's local 512 GB SSD and processors communicate via an 
Infiniband HDR100 fat tree interconnect with 200 Gb/s switches.
%command ibstatus, ibstat, ibv_devinfo ibv_deices
A high-performance Lustre file system provides about 800 TB of storage.
At present, Ookami is running Rocky Linux 8.4, and compilers and 
toolchains from GNU, LLVM, ARM, HPE/Cray, NVIDIA, LLVM, and Fujitsu are installed.

\subsection{User Base}
\begin{figure*}
\centering
    \centering
    \includegraphics[width=0.8\linewidth]{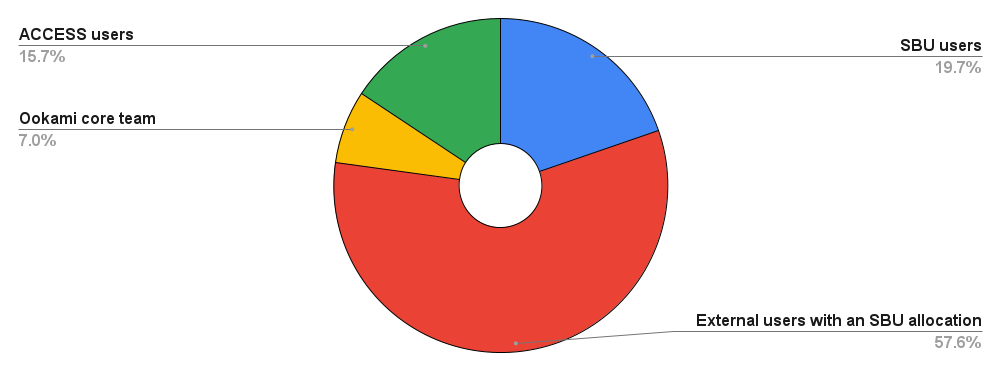}
 \caption{\label{fig:user_pie} Pie chart showing the distribution of users
  on Ookami, 358 as of November, 2023. SBU users are researchers from on-campus. External users are from other academic institutions. The core team is the personnel working on the project, including computational scientists, grad students, and system administrators. ACCESS users are the recent users added via the ACCESS program.}
\end{figure*}

Since Ookami was opened to researchers, 358 users have been onboarded. Those users work on 125 projects, 31 of which have been allocated via ACCESS.
The experience of the user community is very diverse, as they range from undergrad students to professors and professionals with decades of HPC experience.
Figure \ref{fig:user_pie} is a pie
chart illustrating the distribution of users.

% Eva: removing the figure, as I don't think it gives a lot of information and the space can be used better for other content

\subsection{Education and Training}

To cover the complex needs of the user community, the Ookami project has established a multi-modal support approach \cite{siegmanetal2021}.
The extensive documentation, FAQ section, and getting-started guide on the project's website \cite{ookamiurl} allow users to learn everything about the system and its efficient use via self-study. However, because user support is especially important for novel technology, Ookami also offers a Slack channel, to which all new users are automatically added. There they can ask questions, get feedback, or just start an informal conversation with other users as well as with the project team. The average response time is normally within a few minutes, making this the perfect tool for quick help. Office hours, offered twice a week, enable users to ask more detailed questions. These are virtual meetings open to all users, led by Ookami team members. This is the perfect opportunity to do an interactive debugging or profiling session, or to just show applications and get feedback. On average these calls are attended by six to seven persons. Ookami also offers a traditional ticketing system, which is mainly used for installation requests, new user or project requests, or used by users who prefer this approach over the Slack channel.
Regular webinars spanning various topics, e.g.\ profilers, debuggers, and programming languages, are held to support the user community holistically and to provide them with new input for the usage of HPC tools.
The Ookami project team also reaches out to communities, that have potential interest in the cluster, e.g.\ Campus Champions \cite{championsurl}, Science Gateways \cite{sciencegatewaysurl}, and computational researchers with suitable applications, and introduces them to A64FX and the opportunity to use Ookami for their work.

Other outreach activities include an introduction to high-performance computing (HPC) for high school students. Students come to the institute and learn the basics of what a cluster is, and how to use it with the Linux command line. To make the experience more interactive the students visit the server room and work with laptops connected to Ookami to learn about submitting jobs and doing other HPC-related analysis. 
Games for helping to understand fundamental principles in HPC help the students stay motivated. This includes a jigsaw puzzle to explain the concepts of parallelization as well as a mental group exercise to understand how a job scheduler works (Ookami uses the SLURM scheduler). To evaluate the impact of those courses, the students have the chance to fill out a survey before and after the course. The overall response is very positive: the students are excited about HPC, feel more confident in using the Linux environment, and thoroughly enjoy the science nuggets about different research topics that use HPC.

%From my talk: Slack Channel with 337 members, Office Hours, Ticketing system, website with FAQ.Webinars.

%Dave: the transition from education/outreach to describing the physics codes is a bit jarring and makes the paper feel somewhat unfocused.

\section{Performance Studies}

\subsection{V2D}

%Dennis's cluster paper \cite{smolarskietal2022}
%V2D \cite{sm09}

The V2D code 
solves the
Euler equations of inviscid hydrodynamics and multi-species
flux-limited diffusive radiation transport in two spatial
dimensions via finite-difference methods.  It was originally designed for the 
core-collapse supernovae problem of
astrophysics but may be
applied to other radiation-hydrodynamic problems.
Details related to the underlying numerical methods can be
found in \cite{sm09}.  V2D is written in modern Fortran, 
employs MPI for communication, and uses
the the HDF5  library for parallel I/O. 

The problem we chose for our study with V2D
is a the propagation of a 2D Gaussian pulse
of radiation, a diffusive radiation 
transport problem.  The computational
effort is in the solution of a large, sparse,
memory-bandwidth-limited linear system that describes the time
evolution of the radiation distribution. 
The linear system consists of
$x1 \times x2 \times 2$ coupled linear equations, with 
$x1 = 200$, and $x2= 100$ zones in the 
spatial dimensions, and there are 
two radiation species. 
This is a relatively small test problem
chosen to explore the performance 
of SVE optimization. Also, 
we make only limited use of the parallel
capability of V2D when we vary the process topology to adjust
the problem size on each processor. The
linear system is sparse, but it has a
regular structure. The method is
matrix-free, but if the matrix corresponding to this system were stored with a dictionary-like ordering it would form
a banded matrix with five bands. 

We tested combinations of compilers and MPI
implementations. 
As a sample of results, those presented below indicate test runs 
using the GNU (ver.\ 11.1.0), Fujitsu (ver.\
4.5), and HPE/Cray (ver.\ 21.03) compilers.
We compared the CPU times of simulations compiled with
different compilers, both with and without SVE optimization.
The Linux {\tt perf stat} command of the Linux kernel
performance monitor, with the {\tt -e duration\_time} flag, was
used to measure the time of the simulations. This command
measures the entire CPU time of the process. Each configuration
(of the total number of processors used and the process
topology in the $x$1- and $x$2- directions, determining how the
linear system was partitioned) was run several times to confirm
the timing results.

We tested three different compilers on Ookami.
The test problem evolves the radiation energy density for 100 time steps.
Each time step requires the solution of a unique $x_1\times x_2
\times 2$ linear system via the BiCGSTAB algorithm.   Thus this
test problem establishes performance results for the time
evolution of the linear systems for 
of 100 time steps.

All compilers available on Ookami are able to make use of
SVE capabilities.  In our tests, 
the SVE and optimization features were used, but we also
turned off the SVE and other optimization features on some of
the HPE/Cray tests for comparison purposes.
When using a single processor, the executable compiled by GNU 
took the longest to complete the test run, around 363.91 seconds, while
that compiled by the Fujitsu compiler, around 252.31 seconds, 
and the executable compiled using the HPE/Cray compiler (with optimization), 
around 181.26 seconds.  Using more processors, however,
% TONY: need to explain  Does that do it?
%Alan: I think so. Will get confirmed.
with different domain decomposition to CPU core topologies,
the executable compiled by Fujitsu performed
better than the one compiled by HPE/Cray.

The following chart presents these results. The
values in the $N_p$ column refer to
the total number of processors used for the run, with the
numbers in the ``Direction" {\tt NX1} column indicating the
number of domain decomposition tiles in the $x$1 direction and
similarly for the numbers in the {\tt NX2} column. Thus the
product of the two values equals the total number of processors
requested.  The column labeled HPE/Cray (opt) indicates results
obtained with an executable compiled both with both {\tt -O3}
optimization and SVE optimization enabled.  

\begin{center}
\begin{tabular}{|c|r|r|r|c|} \hline
\textbf{$N_p$}&\multicolumn{2}{|c|}{\textbf{Direction}} &
\multicolumn{2}{|c|}{\textbf{Times by Compiler (seconds)}} \\ \hline
      &  \multicolumn{1}{|c|}{\tt NX1} &  \multicolumn{1}{|c|}{\tt NX2}  &  \multicolumn{1}{|c|}{Fujitsu} & \multicolumn{1}{|c|}{HPE/Cray (opt)} \\
 \hline \hline
	40	&	40	&	1	&	  13.97 & 19.12\\ \hline
	40	&	20	&	2	&	  12.96 & 17.37\\ \hline
	40	&	10	&	4	&	  13.04 & 17.16\\ \hline
	50	&	50	&	1	&	  13.05 & 25.56\\ \hline
	50	&	25	&	2	&	  12.09 & 24.07\\ \hline
	50	&	10	&	5	&	  11.40 & 23.51 \\ \hline
\end{tabular}
\end{center}
Further details may be found in \cite{smolarskietal2022}.

\subsection{FLASH}

FLASH is a  simulation software package for addressing multi-scale, multi-physics 
%applications~\cite{Fryxetal00,calder.fryxell.ea:on,flash_evolution,flash_pragmatic,flash_development}. 
applications. 
Initially developed at the University of Chicago
to address  thermonuclear flashes, stellar explosions 
powered by a thermonuclear runaway occurring on the surface or 
in the interiors of compact
stars, FLASH continues to be developed for astrophysics~\cite{townselyetal2019}
and  high-energy-density physics~\cite{flashhed}, and a new code, FLASH-X,
derived from FLASH and with a completely new infrastructure, is under development
and will allow addressing more general problems \cite{Oneal2018}.

At its heart, FLASH is a hydrodynamics plus additional physics (e.g.\
a stellar equation of state) method. FLASH uses the PARAMESH library
to implement adaptive mesh refinement (AMR) to address problems with 
%a wide range of  physical and temporal scales on a block-structured 
a wide range of physical scales on a block-structured 
mesh~\cite{macneice.olson.ea:paramesh,macneice.olson.ea:paramesh*1}.
FLASH is written primarily in modern Fortran and is parallelized primarily
through MPI, although some solvers have been modified to take advantage of threaded
approaches to parallelization~\cite{flash_pragmatic} and
development continues toward a more general design for better thread support~\cite{Daley2012a,CHIUW2019,RADR}.
%allowing threading~\cite{CHIUW2019,RADR}.

PARAMESH manages a block-structured adaptive mesh, with
the data typically in $16 \times 16 \times 16$ zone blocks 
($16 \times 16$ in 2D).
Variables like density, temperature, internal energy, etc., are stored in a 
data container, \texttt{unk}, a Fortran array 
in the form
\texttt{unk(nvar, il\_bnd, iu\_bnd, jl\_bnd, ju\_bnd, kl\_bnd, ku\_bnd, maxblocks)},
where \texttt{nvar} is the number variables, 
\texttt{il\_bnd:iu\_bnd, jl\_bnd:ju\_bnd, kl\_bnd:ku\_bnd} are 
the x, y, and z zone limits, 
and \texttt{maxblocks} is the maximum number of blocks allowed for
a given processor element. Accordingly, block data is typically accessed
block-by-block with the result being that
there is a stride in memory for addressing variables in different zones or
blocks. 
%This feature motivated our interest in investigating the use of
%Huge Pages as a way of improving performance. 

%The modified version of FLASH we use for supernova simulations
%utilizes an advection-diffusion-reaction (ADR) scheme~\cite{VladWeirRyzh06} 
%that propagates reaction progress variables to model the stages of the 
%sub-grid-scale flame. Flame speeds are from the tabulated results
%of previous calculations~\cite{timmes92,Chamulak2007The-Laminar-Fla} and 
%also include enhancement to the burning rate from unresolved buoyancy and 
%background turbulence~\cite{Khok95,townsley.calder.ea:flame,jacketal2014}. 
%Finally we note that the FLASH simulations use double precision
%arithmetic.

As it was one of the marquee applications for the Ookami project, FLASH was ported as soon as
Ookami was up and available. Sorting out the compilers and their options,
versions of MPI, and requisite packages like the HDF5 library took some
effort, but our initial experience with FLASH and other applications
was overwhelmingly positive~\cite{pearc_experiences_2021}.
FLASH ran ``right out of the box" with
several compilers and MPI implementations, scaling reasonably well with no tuning. 
Profiling with Arm MAP~\cite{ArmMap} indicated that thermonuclear 
supernova simulations, our problem of interest,
spent considerable time in the equation of state (EOS) routines 
so we decided to focus our analysis on those routines 
while running 2D supernova simulations. We investigated use
of SVE with the HPE/Cray, Fujitsu, and GNU compilers,
%Is the above true?
%Dennis wrote above that all compilers can use SVE.  
% time dependent answer - at the time of initial FLASH runs, GNU, Fujitsu and Cray delivered SVE. LLVM-13 started SVE support with a -msve-vector-bits flag, and auto-(sve)vectorization was fully enabled starting LLVM 14; NVIDIA compilers have recently started supporting sve instructions (late 22.x or early 23.x). -- smeet 
%Alan: so yes I take it.
but vectorization  
proved difficult due to significant branching in the 
main loops of the EOS routines.
Details of our exploration with scaling results and our 
attempt to utilize the A64FX's SVE instructions and NUMA architecture may be
found in~\cite{flashexperience2022}.

FLASH is capable of full 3D simulations of supernovae, so to 
complement our analysis of the EOS routines we also analyzed 
the hydrodynamics routines. 
We instrumented the code with the Performance Application Programming Interface (PAPI)~\cite{papi}, and found that the number of 
data translation lookaside buffer (DTLB) misses were exceptionally high. 
The DTLB is a special cache that manages the mapping of virtual to physical memory.
Given the stride in memory of the data layout of PARAMESH,
a logical choice was to investigate memory management to study the
high number of DTLB misses and we investigated the use of Huge Pages, blocks of memory larger than the default 64 KB page. 

Huge Pages are a feature that was integrated into the Linux kernel at version 2.6. Memory pages are blocks
of virtual memory managed by the operating system (OS). Huge pages are larger blocks, which means that 
for a given amount of memory, there are fewer pages for the OS to manage. Depending
on the OS, Huge Pages are of, or may be set to, different sizes. Explicitly managing
Huge Pages requires effort and may require changes to an application code, so to ameliorate 
this difficulty, Transparent Huge Pages were implemented in the Linux kernel. In this case,
the kernel is responsible for the creation, management, and use of Huge Pages, making them
an abstraction that is ``transparent" to the user~\cite{THP}. Transparent Huge Pages are by 
default disabled on Ookami and the results presented here are for standard Huge Pages.

For our Huge Pages study, we  ran 2D supernova simulations and 
3D pure hydrodynamics simulations of the Sedov explosion problem~\cite{sedov1959},
one of the standard test problems provided with FLASH. 
We dubbed these ``EOS'' and ``3-d Hydro'' because
the equation of state and three-dimensional hydrodynamics
routines were parts of the code instrumented for performance testing.
We began our study of Huge Pages with the Fujitsu compiler because it readily
made use of Huge Pages with just the addition of the appropriate compiler flags.
Subsequent investigation allowed us to utilize Huge Pages with other 
compilers by linking to the Fujitsu library \texttt{libmpg}.

\begin{figure}[hbtp]
\centerline{\includegraphics[width=\columnwidth]{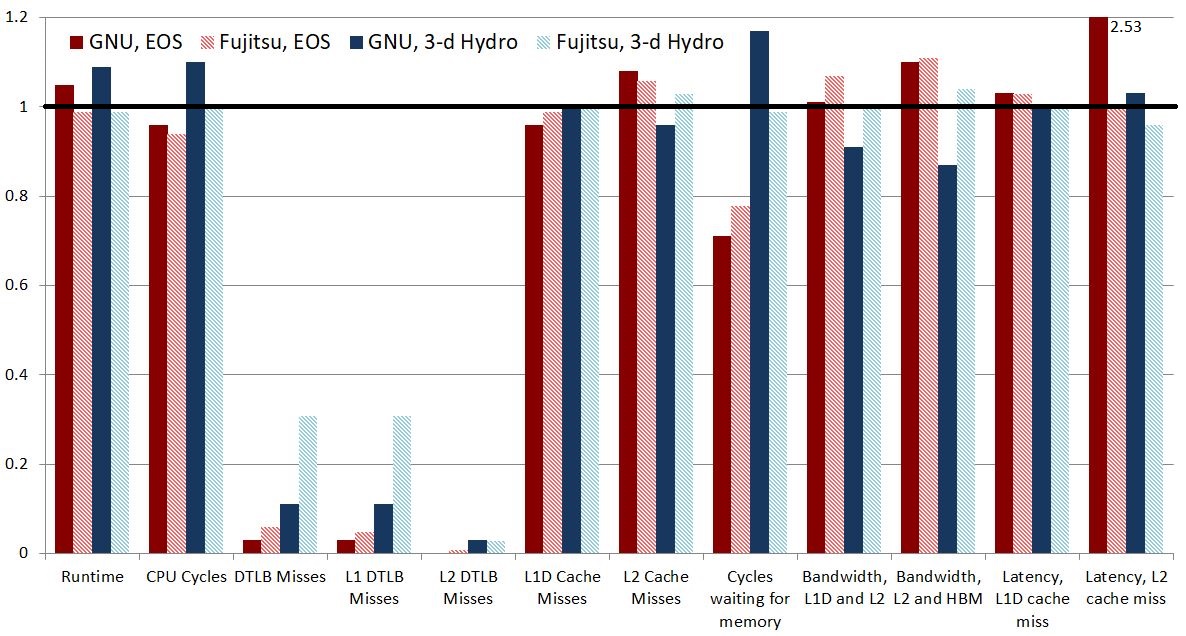}}
\caption{Bar chart showing the ratios of
	performance measures with and without Huge Pages for simulations of the two
 test problems with the Fujitsu and GNU compilers. Each bar is the ratio of each metric for simulations with and without Huge Pages.}
\label{fig:compiler}
\end{figure}

Figure \ref{fig:compiler} shows a comparison between
simulations of the two test problems compiled with
the Fujitsu 4.5 and GCC 12.2.0 compilers. Shown are ratios with and without Huge Pages
of runtime, CPU cycles, DTLB misses, L1 and L2 DTLB misses,
L1D and L2 cache misses, cycles waiting for memory, Bandwidth
between L1D and L2 caches and L2 and HBM, and Latency for
L1D cache misses and L2 cache misses.

\begin{figure}[htbp]
\centerline{\includegraphics[width=\columnwidth]{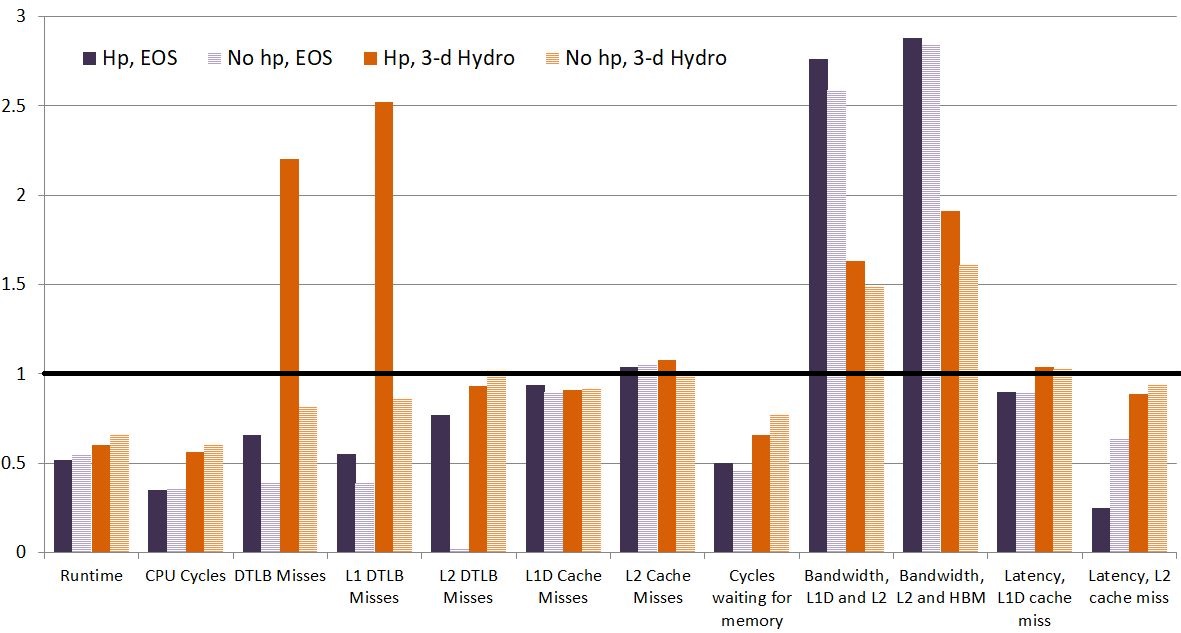}}
\caption{Bar chart showing the ratios of
	performance measures for Fujitsu and GNU compilers for simulations of the two
 test problems with (Hp) and without (No hp) Huge Pages. Each bar is the ratio of each metric for simulations compiled with the Fujitsu to the GNU compiler.}
\label{fig:hp}
\end{figure}

Figure \ref{fig:hp} presents the results of performance tests of 
FLASH for the two problems with and without Huge Pages. Shown are the same performance measures as the compiler test of Figure \ref{fig:compiler}, except the ratios compare the compiler performance of the Fujitsu to the GNU compiler.

The performance results may be summarized as follows:

\begin{itemize}

\item
FLASH performed faster with the Fujitsu compiler, requiring half the runtime and half the number of CPU cycles as with the GNU
compiler.

\item
When FLASH is compiled with the Fujitsu compiler, the 
code spends half as much time waiting for memory than when compiled with the GNU compiler. A significant difference. 

\item
Compiling with the Fujitsu and GNU compilers produced the same number of cache misses.

\item
When compiled with the Fujitsu compiler, FLASH demonstrated twice the memory bandwidth and less latency than with the GNU compiler. 

\item
Use of Huge Pages decreased DTLB misses but did not significantly
decrease the runtime.
\end{itemize}

We concluded that the use of Huge Pages decreased DTLB 
misses as expected, but this decrease did not
affect performance as we expected. We attribute this
finding to the performance of the translation table cache, but further research
is needed to fully understand this result. For FLASH at least,
the Fujitsu compiler produces executables that are twice as fast 
as those produced by the GNU compiler. 
Fujitsu-compiled executables can access HBM much faster, so even though a similar number of cache misses as with the GNU compiler
were seen, misses are not as expensive. But because we 
also found that memory access is only ~20-40\% of the runtime (not illustrated in the plots), we conclude that 
the increased bandwidth can't completely account for the speedup. 
Complete results of our exploration of Huge Pages may be found 
in \cite{flashexperience2022, flash_huge_2022,feldmanetal2023}.

\section{Summary and Conclusions}

Judging from the machine's use, the activity in Office Hours 
and the Slack channel, and the attendance at webinars, we conclude that the project has been 
successful in that it has indeed allowed many users with a
variety of applications to meaningfully explore the Fujitsu
A64FX processor. Our education and outreach efforts have
been effective and have created a community of users that
work well together. 

Our performance studies for the two astrophysics codes produced mixed
results. Both codes were readily ported to Ookami and ran with minimal adjustment. 
The use of SVE improved the performance of V2D but branching in the material EOS routines of FLASH prevented
meaningful vectorization and thus there was no performance improvement.
The use of Huge Pages with FLASH produced a dramatic decrease in DTLB misses but did not improve the
performance of FLASH. This result suggests that DTLB misses do not adversely affect the performance of FLASH.
We attribute this finding to the A64FX's translation table cache, which decreases the latency of virtual 
to physical address translation \cite{a64fxmicro}. 
Finally, our results show that the best performance for V2D and FLASH applications on the A64FX architecture of
Ookami occurs with the use of the Fujitsu compiler.

\ack
Ookami is supported by the US NSF grant \#1927880, and 
this research was supported in part by the US DOE 
under grant DE-FG02-87ER40317. FLASH was developed in part by the US
DOE NSA-ASC and OSC-ASCR-supported Flash Center for Computational 
Science at the University of Chicago.
The authors gratefully acknowledge the generous support 
of the Ookami community. The authors also
thank Jens Domke at RIKEN for very helpful suggestions.

%\section*{References}

%\bibliography{bibliography}

\begin{thebibliography}{10}
\expandafter\ifx\csname url\endcsname\relax
  \def\url#1{{\tt #1}}\fi
\expandafter\ifx\csname urlprefix\endcsname\relax\def\urlprefix{URL }\fi
\providecommand{\eprint}[2][]{\url{#2}}
% Bibliography created with iopart-num v2.0
% /biblio/bibtex/contrib/iopart-num

\bibitem{fugaku_exa}
Matsuoka S 2021 {\em 2021 Symposium on VLSI Circuits\/} pp 1--3

\bibitem{ookamiurl}
IACS 2020 Ookami homepage
  \urlprefix\url{https://www.stonybrook.edu/commcms/ookami/}

\bibitem{pearc_experiences_2021}
Burford A, Calder A, Carlson D, Chapman B, Coskun F, Curtis T, Feldman C,
  Harrison R, Kang Y, Michalowicz B, Raut E, Siegmann E, Wood D, DeLeon R,
  Jones M, Simakov N, White J and Oryspayev D 2021 {\em Practice and Experience
  in Advanced Research Computing\/} PEARC '21 (New York, NY, USA: Association
  for Computing Machinery) ISBN 9781450382922
  \urlprefix\url{https://doi.org/10.1145/3437359.3465578}

\bibitem{sm09}
{Swesty} F~D and {Myra} E~S 2009 {\em \apjs\/} {\bf 181} 1--52

\bibitem{Fryxetal00}
{Fryxell} B, {Olson} K, {Ricker} P, {Timmes} F~X, {Zingale} M, {Lamb} D~Q,
  {MacNeice} P, {Rosner} R, {Truran} J~W and {Tufo} H 2000 {\em The
  Astrophysical Journal Supplement Series\/} {\bf 131} 273--334

\bibitem{calder.fryxell.ea:on}
{Calder} A~C, {Fryxell} B, {Plewa} T, {Rosner} R, {Dursi} L~J, {Weirs} V~G,
  {Dupont} T, {Robey} H~F, {Kane} J~O, {Remington} B~A, {Drake} R~P, {Dimonte}
  G, {Zingale} M, {Timmes} F~X, {Olson} K, {Ricker} P, {MacNeice} P and {Tufo}
  H~M 2002 {\em The Astrophysical Journal Supplement Series\/} {\bf 143}
  201--229

\bibitem{flash_development}
{Dubey} A, {Antypas} K, {Calder} A, {Fryxell} B, {Lamb} D, {Ricker} P, {Reid}
  L, {Riley} K, {Rosner} R, {Siegel} A, {Timmes} F, {Vladimirova} N and {Weide}
  K 2013 {\em 2013 5th International Workshop on Software Engineering for
  Computational Science and Engineering (SE-CSE)\/} pp 1--8

\bibitem{nsfaccess}
 2023 {NSF} {ACCESS} program \urlprefix\url{https://access-ci.org}

\bibitem{armv8.2}
Mann B 2017 Arm architecture - armv8.2-a evolution and delivery
  \urlprefix\url{https://community.arm.com/arm-community-blogs/b/architectures-and-processors-blog/posts/arm-architecture-armv8-2-a-evolution-and-delivery}

\bibitem{siegmanetal2021}
Siegmann E, Calder A, Feldman C and Harrison R~J 2021 {\em 2021 IEEE/ACM Ninth
  Workshop on Education for High Performance Computing (EduHPC)\/} pp 16--23

\bibitem{championsurl}
 2023 Campus {Champions}
  \urlprefix\url{https://campuschampions.cyberinfrastructure.org/}

\bibitem{sciencegatewaysurl}
 2023 Science {Gateways} \urlprefix\url{https://sciencegateways.org/}

\bibitem{smolarskietal2022}
Smolarski D~C, Swesty F and Calder A~C 2022 {\em 2022 IEEE International
  Conference on Cluster Computing (CLUSTER)\/} (Los Alamitos, CA, USA: IEEE
  Computer Society) pp 545--548
  \urlprefix\url{https://doi.ieeecomputersociety.org/10.1109/CLUSTER51413.2022.00071}

\bibitem{townselyetal2019}
{Townsley} D~M, {Calder} A~C and {Miles} B~J 2019 {\em Journal of Physics
  Conference Series\/} ({\em Journal of Physics Conference Series\/} vol 1225)
  p 012004 (\textit{Preprint} \eprint{1908.06176})

\bibitem{flashhed}
 2023 {The Flash Center for Computational Science}
  \urlprefix\url{https://flash.rochester.edu}

\bibitem{Oneal2018}
O'Neal J, Weide K and Dubey A 2018 {\em WSSSPE6.1, colocated with eScience
  2018, Amsterdam, Netherlands\/}

\bibitem{macneice.olson.ea:paramesh}
{MacNeice} P, {Olson} K~M~{Mobarry} C, {de Fainchtein} R and {Packer} C 1999
  {\em NASA Tech. Rep. CR-1999-209483\/}

\bibitem{macneice.olson.ea:paramesh*1}
{MacNeice} P, {Olson} K~M~{Mobarry} C, {de Fainchtein} R and {Packer} C 2000
  {\em Comput. Phys. Commun.\/} {\bf 126} 330--354

\bibitem{flash_pragmatic}
Dubey A, Calder A, Fisher R, Graziani C, Jordan G, Lamb D, Reid L, Townsley D
  and Weide K 2013 {\em International Journal of High Performance Computing
  Applications\/} {\bf 27} 360--373 ISSN 1094-3420

\bibitem{Daley2012a}
Daley C, Bachan J, Couch S, Dubey A, Fatenejad M, Gallagher B, Lee D and Weide
  K 2012 {\em {TACC}-Intel Highly Parallel Computing Symposium\/} poster

\bibitem{CHIUW2019}
Dubey A 2019 Programming abstractions for orchestration of hpc scientific
  computing \url{https://chapel-lang.org/CHIUW2019.html} keynote, Chapel User's
  Group Meeting

\bibitem{RADR}
Dubey A 2019 Dynamic resource management, an application perspective
  \url{https://project.inria.fr/resourcearbitration/program/} invited talk,
  RADR, co-located with IPDPS

\bibitem{ArmMap}
ARM 2021 Arm forge/map
  \urlprefix\url{https://www.arm.com/products/development-tools/server-and-hpc/forge/map}

\bibitem{flashexperience2022}
Feldman C, Michalowicz B, Siegmann E, Curtis T, Calder A and Harrison R 2022
  {\em International Conference on High Performance Computing in Asia-Pacific
  Region Workshops\/} HPCAsia 2022 Workshop (New York, NY, USA: Association for
  Computing Machinery) p 72–77 ISBN 9781450395649
  \urlprefix\url{https://doi.org/10.1145/3503470.3503478}

\bibitem{papi}
 2022 {Performance Application Programming Interface}
  \urlprefix\url{http://icl.cs.utk.edu/papi/}

\bibitem{THP}
{Red Hat} 2022 {5.2. Huge Pages and Transparent Huge Pages}
  \urlprefix\url{https://access.redhat.com/documentation/en-us/red_hat_enterprise_linux/6/html/performance_tuning_guide/s-memory-transhuge}

\bibitem{sedov1959}
{Sedov} L~I 1959 {\em {Similarity and Dimensional Methods in Mechanics}\/}

\bibitem{flash_huge_2022}
Calder A~C, Feldman C, Siegmann E, Dey J, Curtis A, Chheda S and Harrison R~J
  2022 {\em 2022 IEEE International Conference on Cluster Computing
  (CLUSTER)\/} (Los Alamitos, CA, USA: IEEE Computer Society) pp 539--544
  \urlprefix\url{https://doi.ieeecomputersociety.org/10.1109/CLUSTER51413.2022.00070}

\bibitem{feldmanetal2023}
Feldman C, Chheda S, Calder A, Siegmann E, Dey J, Curtis T and Harrison R 2023
  {\em Practice and Experience in Advanced Research Computing\/} PEARC '23 (New
  York, NY, USA: Association for Computing Machinery) ISBN 9781450382922
  \urlprefix\url{https://doi.org/10.1145/3569951.3597583}

\bibitem{a64fxmicro}
Fujitsu 2023 A64fx microarchitecture manual
  \urlprefix\url{https://github.com/fujitsu/A64FX/blob/master/doc/A64FX_Microarchitecture_Manual_en_1.3.pdf}

\end{thebibliography}

\providecommand{\newblock}{}

\end{document}